\documentclass[authoryear,preprint,review,12pt]{elsarticle}

\usepackage{graphicx}

\usepackage{amssymb}
\usepackage[utf8]{inputenc} % Use UTF-8 encoding
\usepackage[T1]{fontenc}    % Use T1 font encoding for proper font support

\usepackage{url}

\journal{Remote Sensing of Environment}

\begin{document}

\begin{frontmatter}



\title{WHALES: an optimized retracker for satellite radar altimeter waveforms in sea state applications}


\author[dgfi]{Marcello Passaro}

\author[CNRS]{Guillaume Dodet}

\author[CNRS]{Fabrice Ardhuin}

\author[ESA]{Paolo Cipollini}

\affiliation[dgfi]{organization={Deutsches Geodätisches Forschungsinstitut, Technical University of Munich (DGFI-TUM)},%Department and Organization
            addressline={Arcisstraße 21}, 
            city={Munich},
            postcode={80333}, 
            country={Germany}}

\affiliation[CNRS]{organization={Univ. Brest, Ifremer, CNRS, IRD},%Department and Organization
            addressline={LOPS}, 
            city={Plouzan\'{e}},
            postcode={29280}, 
            country={France}}

\affiliation[ESA]{organization={European Space Agency},%Department and Organization
            addressline={ESTEC}, 
            city={Noordwijk},
            postcode={2201}, 
            country={The Netherlands}}

\begin{abstract}
The latest version of the European Space Agency's Sea State Climate Change Initiative database adopts a dedicated algorithm (retracker) to reprocess two decades of satellite altimetry measurements and provide long time series of significant wave height in the global ocean. This paper describes the main characteristics of this algorithm, called WHALES, and analyzes the impact of algorithm choices on measurement physics, particularly the weighted analysis of residuals in the cost function. Moreover, the impact of WHALES on the sea state database is analyzed in terms of noise reduction and scatter index with in situ data, with a particular focus on the coastal zone, where WHALES primarily improves data quality and quantity compared to previous approaches. We found that valid data records increased by 30\% at 5 km from the coast after applying the retracker discussed here.
\end{abstract}


\begin{highlights}
\item The WHALES algorithm estimates wave height from radar altimetry signals. 
\item The impact of algorithm choices on measurement physics is analyzed.
\item Validation shows noise reduction and improved data quality in coastal zones.  
\end{highlights}

\begin{keyword}
satellite altimetry \sep significant wave height \sep retracking



\end{keyword}

\end{frontmatter}


\section{Introduction}
\label{introduction}
Sea state information is important for the analysis of air-sea interactions \citep{Cronin&al.2019} and key to all human activities at sea and on the coast. The design of any marine or coastal structure particularly requires long term time series to establish the wave climate \citep{Battjes1984}. In this context, the extrapolation from in situ time series has been largely complemented by numerical wave models, and satellite measurements are key for the calibration and validation of these models in all conditions. The present paper provides a description and performance analysis of the retracking method that is used to estimate the significant wave height from the delay-only processing of radar altimetry signals (also called Low Resolution Mode processing, or LRM) in the European Space Agency's (ESA) Sea State  Climate Change Initiative (CCI) databases \citep[e.g.][]{Dodet&al.2020}. This method is called 'WHALES' and is an evolution of the Adaptive Leading Edge Subwaveform (ALES) method that was developed for coastal altimetry \citep{Passaro&al.2014}, with a particular focus on retrieving wave heights in both open ocean and coastal environments. 

\begin{figure*}[!htb]
\centerline{\includegraphics[width=\linewidth]{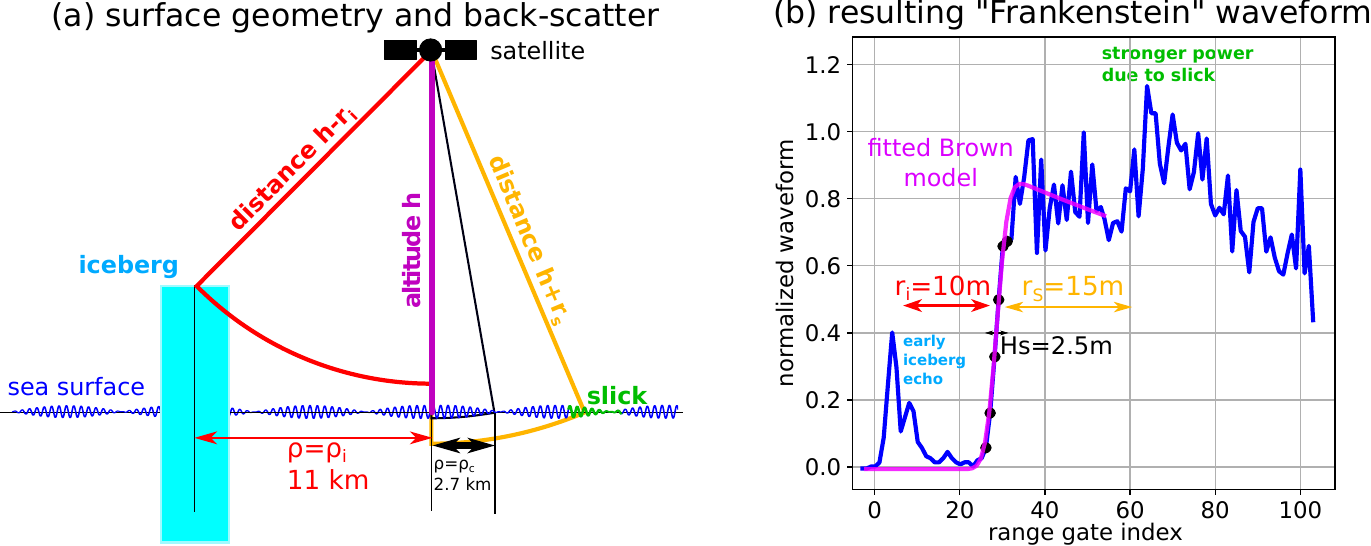}}
\caption{
(a) surface geometry schematic for a scenario including an iceberg and a surface slick and (b) a possible resulting altimeter waveform for a Jason altimeter. In the schematic the vertical scale of the surface is exaggerated by a factor $\alpha=90$ compared to the horizontal, and the satellite altitude is reduced by $\alpha$, so that  the Chelton et al. radius $\rho_c$ is properly scaled, but the incidence angles are exaggerated. The leading edge is defined by a few range gates (black dots) for this $H_s=2.5$~m case. A much larger $\sigma_0$ at larger ranges $h+r_s$ creates a higher waveform peak for range gate indices 60-80. This very unlikely "Frankenstein" waveform was created by combining 2 waveforms from Jason-2, that are further discussed in section \ref{discussion_adjustments}. 
%vertical surface stretch: 3000/20: 159 .   altitude : 15*150 . 
}
\label{fig:schematic}
\end{figure*}

With continuous measurements since the launch of ERS-1 in 1991, satellite radar altimeters provide the backbone of the Sea State CCI databases. %, which are routinely updated with data from operational satellites. %, including the Copernicus Sentinel 3 and 6 series. 
These altimeters primarily measure the power of echoes as a function of the round trip time between the transmission and reception of radar pulses. This power as a function of time is called a waveform, and the time resolution $\Delta t$, given by the inverse of the radar pulse bandwidth, translates to a range resolution $\delta_r = c \Delta t /2$ which is 47 cm for the 320-MHz bandwidth used in all current Ku-band spaceborne altimeters. Waveforms are typically discretized in about 128 range gates, and averaged over about 100 pulses. For any range gate, the received power combines echoes from all possible targets at the same distance from the radar (slant range, illustrated in Fig. \ref{fig:schematic} for different targets).  This in turn means that for a conventional altimeter (one that does not use doppler or interferometric processing) the vertical displacement of targets cannot be separated from their horizontal displacement. For an island or iceberg of height $z$ at a distance $\rho$ from nadir, and using a flat surface approximation, which is justified as the satellite altitude $h$ is much larger than the radar footprint radius, the square range of the echo is $(h+r)^2 = (h-z)^2 +\rho^2$ and using $h \gg z$  we get $r\approx\rho^2/2h -z$, i.e. the expression for the range anomaly $r$ (difference between range to the target and altitude of the satellite). In the ocean away from coasts and sea ice, the surface elevations observed by the radar in its footprint are dominated by wind-generated waves. Assuming, as a first approximation, that the wave elevations have a Gaussian distribution of standard deviation $\sigma_h$, the significant wave height (which is defined as the average crest-to-trough height of the highest 1/3 of the waves) is $H_s=4 \sigma_h$. As a result, the radius of the   "oceanographic footprint" defined by  \cite{Chelton&al.1989} as 
\begin{equation}
    \rho_c = \sqrt{2 H_s h},
\end{equation} is the distance from nadir where a sea surface higher than $H_s/2$ (which has a probability of only 2\%) has a range larger than $h+H_s/2$. This "oceanic footprint" contains the vast majority of the ocean scatterers that contribute to the leading edge, defined as ranges between $h-H_s/2$ and $h+H_s/2$ (black dots in Fig. \ref{fig:schematic}.b).

%%%%%%%%%%%%%%%%%%%%%%%%%%%%%%%%%%%%%%%%%%%%%%%%%%%%%%%%%%%%%%%%%%%%%%%%%%%%%%%%%%
%%%%%%%%%%%%%%%%%%%%%%%%%%%%%%%%%%%%%%%%%%%%%%%%%%%%%%%%%%%%%%%%%%%%%%%%%%%%%%%%%%

The retracking procedure consists of fitting the measured waveform with a theoretical waveform that is a function of the epoch $\tau$ (the time delay between the radar pulse being transmitted and the point in the waveform that corresponds to the local mean sea surface elevation), significant wave height $SWH$ (defined as 4 times the local standard deviation of the surface elevation) and a local normalized radar cross section (NRCS) $\sigma_0$. The $SWH$ estimated in the retracking procedure is different from the theoretical $H_s$, because it includes the effect of the wave groups, which breaks the hypothesis of the surface elevations in the footprint being Gaussian distributed. We will therefore use both notations depending on the use throughout this study. 
The reflectivity $\sigma_0$ is, for a water surface, inversely proportional to the surface slope variance \citep{Vandemark&al.2004}.   

When referring to "local" variables, we expect these quantities to vary little at the scale of at least $\rho_c$. However, $\rho_c$ is generally much smaller than the area illuminated by the radar which is defined by the antenna pattern and the satellite altitude, with a diameter $\rho_r$ of the order of 30 km for the Jason series. Perturbations of $\sigma_0$ at distances from nadir between $\rho_c$ and $\rho_r$ give perturbations of the trailing edge, as shown in Fig. \ref{fig:schematic}. These can be caused by rain, and may look like the effect of mispointing \citep{Quartly&al.1996}, or areas of low winds, with the added effect of surface slicks \citep{Tournadre&al.2006}.

The simplest theoretical waveform was given by \cite{Brown1977}, and it takes into account the radar parameters (beam width and mispointing angle) and satellite altitude, and assumes that all three geophysical parameters  $\tau$,  $H_s$ and $\sigma_0$ are homogeneous over the radar footprint and do not depend on the distance from nadir. %, with the same Gaussian distribution of the elevation of scattering facets at the sea surface at each range.  
The Brown waveform,  a good model for usual open ocean echoes, is characterised by a rising leading edge that becomes less steep as $H_s$ increases, and a slowly decreasing trailing edge. In practice the waveform is perturbed by Rayleigh fading which introduces speckle noise, and non-homogeneities of surface level, wave properties and $\sigma_0$ within the radar footprint. These perturbations are most severe and frequent in coastal areas and near sea ice. However, they also occur in the open ocean, due to scenarios like those schematized in Fig. \ref{fig:schematic}. A general processing algorithm should ideally account for all these possibilities, either at the retracking stage or at a later editing stage that may remove outliers in the retracked parameters. The advent of the ALES retracker has already shown the way to keep the quality of the retrievals in the open ocean while improving the data quality and quantity in the 25 km near the coast, where waveforms tend to be  corrupted by heterogeneous backscattering from land and sheltered water. Nevertheless, besides still using the full echo to retrieve parameters that are located on the leading edge, the standard retracking method is still affected by a suboptimal distribution of the residuals in the fitting process, which results in high level of noise in the estimations. This motivated the development of Maximum Likelhood estimators that use a logarithmic cost function which uses best the lower part of the waveform when perturbations are dominated by a multiplicative noise like speckle noise \citep{Challenor&Srokosz1989,Gomez-Enri&al.2007,tourain2021}. Also, we now understand that the SWH produced by standard retracking procedure is equivalent to a convolution of the map of local wave heights by a smoothing kernel $J_H$ \citep{DeCarlo&Ardhuin2024}. Keeping the full waveform with constant weights gives a broad kernel $J_H$, with a relatively coarse resolution. We know that changing the weights or the cost function also changes the shape of $J_H$, which allows us to optimise the trade-off between the desired resolution and the precision of the estimations \citep{DeCarlo&Ardhuin2024}. 

WHALES is designed as a unified way to solve these and other problems currently affecting the standard product. The general goal of WHALES is to retrieve as many valid data as possible from the waveform, with the lowest noise level, so that we may use not just the 1Hz median values of the 20Hz data, but also we may produce meaningful estimates of the statistics of 20Hz data that can be useful to reveal coastal processes \citep{Passaro&al.2021}, ocean wave group properties and wave-current interactions \citep{DeCarlo&al.2023}. WHALES is based on two principles:

1. A subwaveform strategy (inherited from the ALES retracker) to focus the retracking on the portion of the signal of interest, avoiding heterogeneous backscattering in the trailing edge. This yields efficiency in the coastal zone and a better representation of the
oceanic scales of variability.

2. The application of a weighted fitting solution, whose weights depend on SWH in order to guarantee a more uniform distribution of the residuals during the iterative fitting. This guarantees significantly more precise estimations, and also contributes to a higher effective resolution. 

The details of the WHALES algorithm are presented in section \ref{methodology}. A validation of the Sea State CCI database based on WHALES follows in section 3. Section 4 analyses the effective footprint shape from a theoretical point of view and using simulated waveforms. Possible trade-offs and adjustments are discussed in section 5 with some illustrations using Jason-2 data. 

\section{Methodology}
\label{methodology}

\subsection{WHALES retracker}

WHALES is a two-pass retracker. Although it could be combined with any theoretical waveform, we have chosen the simple \cite{Brown1977} waveform (as described in the formulation of section 3.1 in \cite{passaro2014}) and we fit the three geophysical parameters $\tau$,  $SWH$ (which is linked to the rise time $\sigma_c$ of the leading edge of the waveform) and $\sigma_0$ (which is linked to the amplitude of the signal $P_u$). The altimeter measurements are discretized in elements called “gates”. In WHALES, the first gate number is identified with the index 0. The x-axis of a waveform is sampled in time, or more exactly the delay time between transmission and reception of a radar pulse. For example for Jason-3:

$x = \{0, 1 \cdot \Delta t, 2 \cdot \Delta t, \ldots, 103 \cdot \Delta t\}$

Where $\Delta t$ is the spacing between two consecutive gates in time (3.125 ns in Jason-3, i.e. the time resolution as defined in section \ref{introduction}).

Before the fitting, the leading edge identification, which includes also the normalisation of the waveform power, is performed following these substeps:

\begin{enumerate}
    \item The waveform is normalised with normalisation factor \( N \), where \( N = 1.3 \cdot \mathrm{median}(\mathrm{waveform}) \).
    \item The leading edge starts when the normalised waveform has a rise of 0.01 units compared to the previous gate (startgate).
    \item At this point, the leading edge is considered valid if, for at least four gates after startgate, it does not decrease below 0.1 units (10\% of the normalised power).
    \item The end of the leading edge (stopgate) is fixed at the first gate $i$ in which the difference of the normalised power between gate $i+1$ and $i$ is negative (i.e., the signal starts decreasing and the trailing edge begins), unless the signal keeps growing in the following 3 gates. 
    % This was problematic for SARAL, it is also problematic for Hs > 10 m. (to be discussed below). Instead or in complement of the smoothing introduced by Marine, I suggest to use a threshold value (0.6 or 0.8) above which the leading edge must rise before the end is detected. 
\end{enumerate}

We note that the thresholds used in the steps for leading edge identification were determined through trial and error, guided by empirical experience with both synthetic and real waveforms. They are presented here to document the choices made in the current version of the algorithm. However, minor adjustments to these thresholds do not significantly affect the results. 
In particular, the scope of the normalisation is indeed to take as reference power a value close to the
maximum of the leading edge and, in the case of oceanic waveforms with standard trailing
edge noise, the proposed factor \( N \) is a good compromise, as shown in Figure \ref{additionalfigures}.

The first pass of WHALES involves a subwaveform that goes from startgate to stopgate+1. It
is therefore a leading-edge-only subwaveform retracking. The vector of weights $w_i$ is filled with 1s. The convergence is then found iteratively by means of a Nelder-Mead estimator \citep{nelder1965} applied to a least square cost function: 

\begin{equation}
F(\theta) = \sum_{i} w_i \left( y_i - \hat{y}_i \right)^2
\end{equation}

where \( y_i \) are the values of the real waveforms, and \( \hat{y}_i \) are the the values fitted with the chosen waveform model (Brown), \( \theta \) represents the set of parameters (unknowns) being optimized. In case convergence is not reached, a new attempt is
performed extending the subwaveform, until convergence or until the waveform limit.\footnote{Note that initial conditions have to be assigned to each unknown. The initial conditions applied in this study are: 
\[
\tau_0 = \mathrm{startgate} - 1 \quad 
\sigma_{c_0} = \frac{\mathrm{stopgate} - \mathrm{startgate}}{2\sqrt{2}} \quad 
P_{u_0} = 2 \times \mathrm{mean}(D[\mathrm{startgate} : \mathrm{stopgate}])
\]
Where \( D \) is the normalised waveform.}

After the first pass, the WHALES coefficients are applied to extend the subwaveform. As
explained in the next section, the issue is one of defining an appropriate new stopgate for
the second pass retracking based upon the SWH estimates from the first pass. % Note that SWH cannot exceed 15 m for this equation ... 
For Jason-3, the following coefficients are used:
\begin{equation}
\label{linearrelationship}
\mathrm{Stopgate} = \lceil \mathrm{Tracking\ point} + 3.89 + 3.86 \times \mathrm{SWH} \rceil
\end{equation}
These coefficients were recomputed specifically for the current purposes as explained in section \ref{WHALEScoefficients}.

Using the new limits of the subwaveform, a second fit is performed using the
same initial conditions of the first pass. This time, the SWH estimation of the first pass is also used to identify the proper set of weights (see section \ref{WHALESweights}). WHALES therefore is adaptive in both the subwaveform width and the weights.

The SWH estimated in the second pass is instrumentally corrected by means of a look-up
table that takes into account the bias due to the Gaussian approximation of the point target response in the Brown model. The estimated power is converted in dB and corrected by atmospheric correction and scaling factor, whose fields are contained in the Geophysical Data Records (GDRs) of each mission. This constitutes the output of the algorithm (backscatter coefficient). 
%The epoch is not %\(\sigma_0\)
%provided in the output since its precision and accuracy has not been verified.

The “Fitting Error on the leading edge” (Err) is used as a quality measure for the fitting. It is computed as the RMS difference between the fitted and the measured waveform, considering only the gates of the leading edge. When \(\mathrm{Err} > 0.3\), the quality flag is set to 1, i.e., the quality of the fitting is bad.

\subsection{WHALES coefficients}
\label{WHALEScoefficients}

The key consideration leading to WHALES as a two-step retracker is that a leading-edge only retracker, although also providing results for waveforms that do not conform to the Brown model, has worse noise performances than a full-waveform retracker and therefore would not guarantee the homogeneity of the result. For best accuracy, the subwaveform width for the second pass
must be optimized such that, while it includes the whole of the leading edge, it also exploits part of the information in the trailing edge for as much as is necessary to bring the noise below a desired threshold. This compromise is aimed at avoiding that artifacts such as bright targets or land intrusion in the trailing edge may prevent the Brown model from accurately describing the shape.  

The relationship between SWH and stopgate was derived from Monte Carlo simulations. For each value of SWH ranging from 0.5 to 10 m in steps of 0.5 m, 10,000 waveforms were simulated with the Brown model adding realistic Rayleigh noise (a multiplicative speckle noise with an intensity defined by the number of averaged pulses), and then averaged to create a simulated high-rate waveform. The resulting waveforms were retracked over the entire waveform, and then over sub-waveform windows with startgate = 1 and variable stopgate, and the RMS errors (RMSE) of the SWH were computed (instead of the RMSE for $\tau$, as in ALES).

\begin{figure}[!htb]
  \centering
  \includegraphics[width=\textwidth]{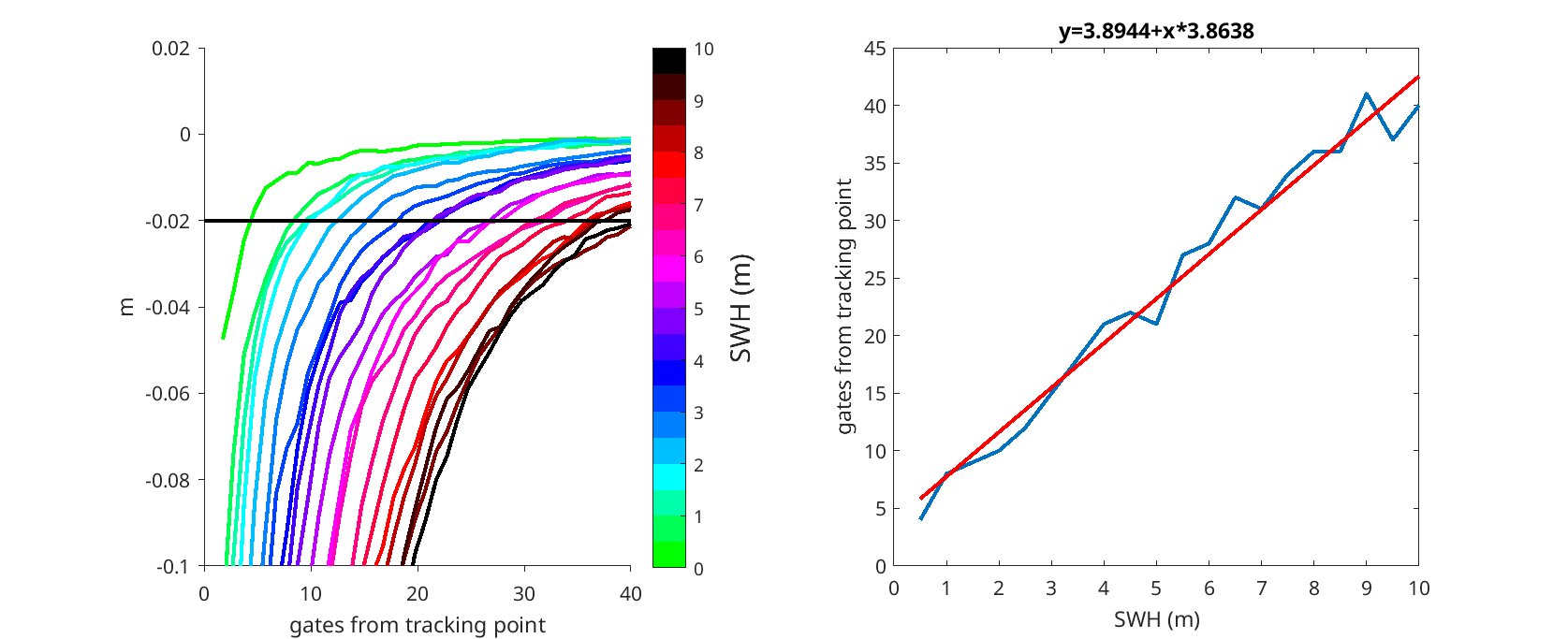}
  \caption{(Left) Difference of the RMSEs between the "full waveform" estimate and the
subwaveform estimates as a function of the stopgate position. (Right): linear relationship
obtained by setting a tolerance in the RMSE difference of the SWH (black bar in the left plot). }
  \label{montecarlo}
\end{figure}

The difference of the RMSEs between the "full waveform" estimate and the subwaveform
estimates is displayed as a function of the stopgate position in the figure below \ref{montecarlo}. The x-axis corresponds to the width of the sub-waveform, expressed as the number of gates from the tracking point to the stopgate. The results for each SWH level are coded in different colors. For all three parameters, the curves converge asymptotically to the full waveform estimates, as expected for this idealized case of "pure-Brown" response of the ocean surface. The relation needed for step \ref{linearrelationship} of WHALES is obtained by setting a tolerance in the RMSE difference of the SWH. In order for WHALES to optimize the need to retrieve signals whose trailing edge is corrupted, the tolerance bar was set to 2 cm at 20 Hz, i.e., 0.45 cm at 1 Hz.

\subsection{WHALES weights}
\label{WHALESweights}

To derive the adaptive set of weights that is used to find convergence in WHALES, a second Monte Carlo simulation was performed. The objective was to base the weighting on the uncertainty of the fitting along the leading edge of the waveform. As an estimation of the uncertainty, the standard deviation (std) between the simulated echo and a large number of fitted waveforms was used.

\begin{figure}[!htb]
  \centering
  \includegraphics[width=\textwidth]{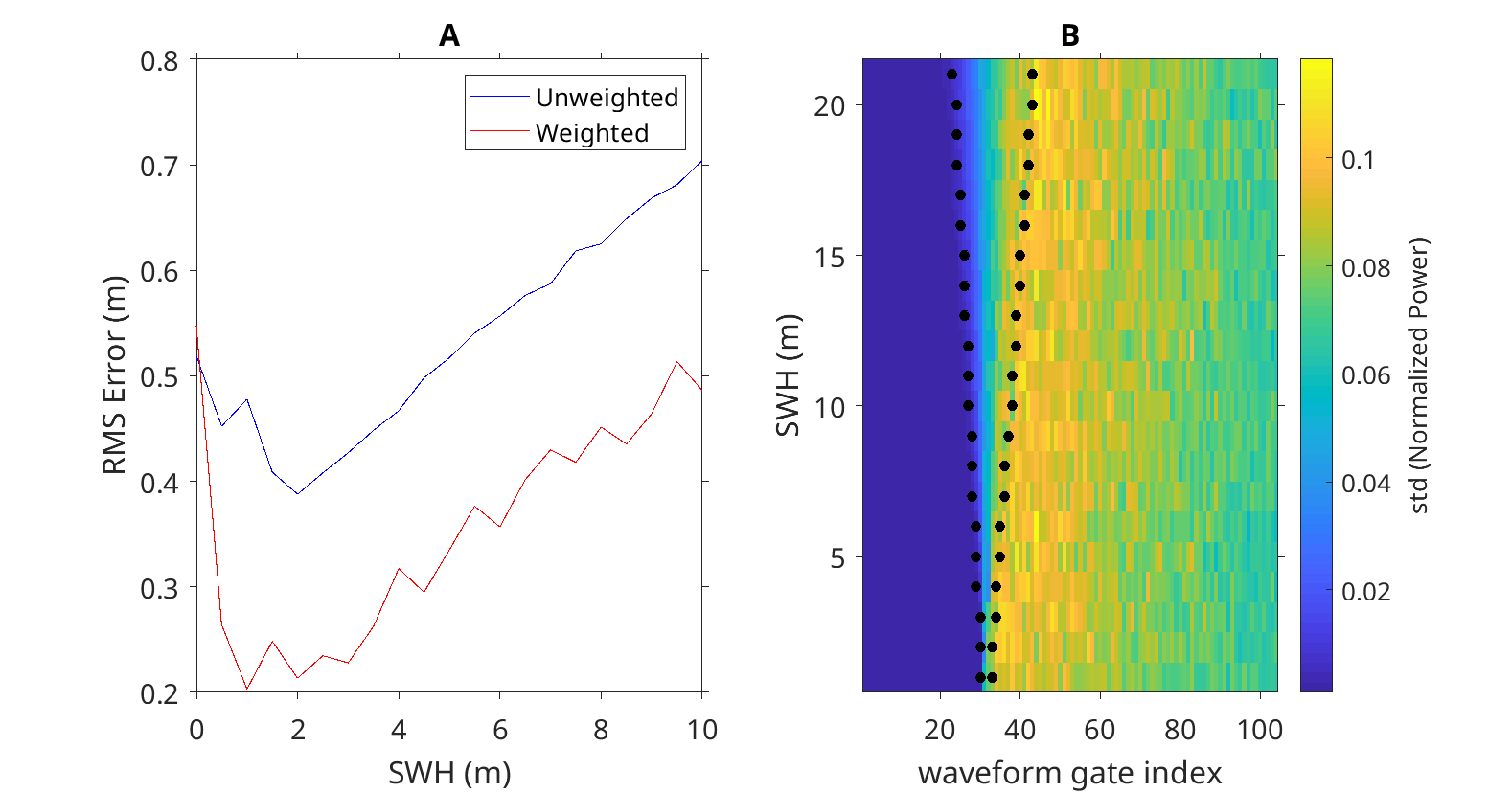}
  \caption{(A): RMS of SWH estimates using unweighted (blue) and weighted (right) residuals in the retracking applied on synthetic waveforms generated for difference SWH values; (B): std of the difference between the unweighted retracking and the simulated signal at each waveform gate, as a result of the Montecarlo simulation described in this study. Results are grouped on the y-axis according to the SWH of the simulated signal. The black dots correspond to the start and the end of the leading edge.}
  \label{weights_figure}
\end{figure}

Simulated waveforms were again generated according to the Brown model as previously reported (SWH at steps of 0.5 m from 0 to 10 m; 10,000 waveforms per SWH value). Each waveform was retracked through the unweighted Nelder-Mead estimator. For each SWH value we stored:
\begin{itemize}
    \item The value of the residuals between the simulated echo and the fitted waveform
    \item The position of the start and the end of the leading edge 
\end{itemize}
For each SWH value, there are therefore 10,000 values of residuals at each waveform gate, and their std can be computed. The std of the residuals for varying SWH is shown in figure \ref{weights_figure}B. It is noted that the std is low in the initial part of the leading edge and increases with the increasing waveform power. The increasing width of the leading edge as the SWH grows is also visible. The weights chosen in the WHALES retracking are the inverse of this std, i.e., the so-called “Statistical Weighting”, which in statistics is a recommended choice when the uncertainties of the different points to fit are very different from each other \citep{wolberg2006}. 
Figure \ref{weights_figure}A shows the effect of the application of the weights in the retracking process, by plotting the SWH Root Mean Square Error (RMSE) in the case of unweighted (blue curve) and weighted (red curve) least square estimations. A notable improvement of factor 1.5 to 2 is obtained at all generated SWH values, with the exclusion of SWH below 1 m, which cannot be correctly resolved in satellite altimetry due to the poor sampling of the leading edge \citep{smith2015}.

%HERE MISSING: DESCRIPTION OF rms_simulation_figure

\section{Validation}
\label{validation}

The 20Hz SWH estimations from the WHALES retracker were already validated in a round-robin exercise against a large number of algorithms in the context of the ESA Sea State CCI project \citep{schlembach2020}. The WHALES retracker was then chosen as the algorithm of choice for the so-called "Low Resolution Mode" (LRM) data, i.e. all missions except Sentinel-3a in the current version of the Sea State CCI database (version3). The data were averaged at 1 Hz and cross-calibrated following the procedure described in \cite{piolle2025seastate}. 

%Firstly, 20Hz measurements located on land are excluded, as well as SWH values smaller than -0.5m or larger than 30m. Subsequently, remaining outliers are excluded using a 3-sigma criterion based on the maximum absolute deviation. Finally, the 1Hz SWH value is computed as the median of the remaining 20Hz SWH values.

The objective of this section is to evaluate the impact that the introduction of WHALES has had on the database with a particular focus on the coastal data. This is done in \ref{impact_noise} comparing version3 with a previous version of the database which does not contain WHALES estimations (version1) and in \ref{impact_insitu} comparing alternative solutions to WHALES against in situ records using the Jason-3 mission data.

\subsection{Impact of WHALES on noise level}
\label{impact_noise}
\begin{figure}[!htb]
  \centering
  \includegraphics[width=\textwidth]{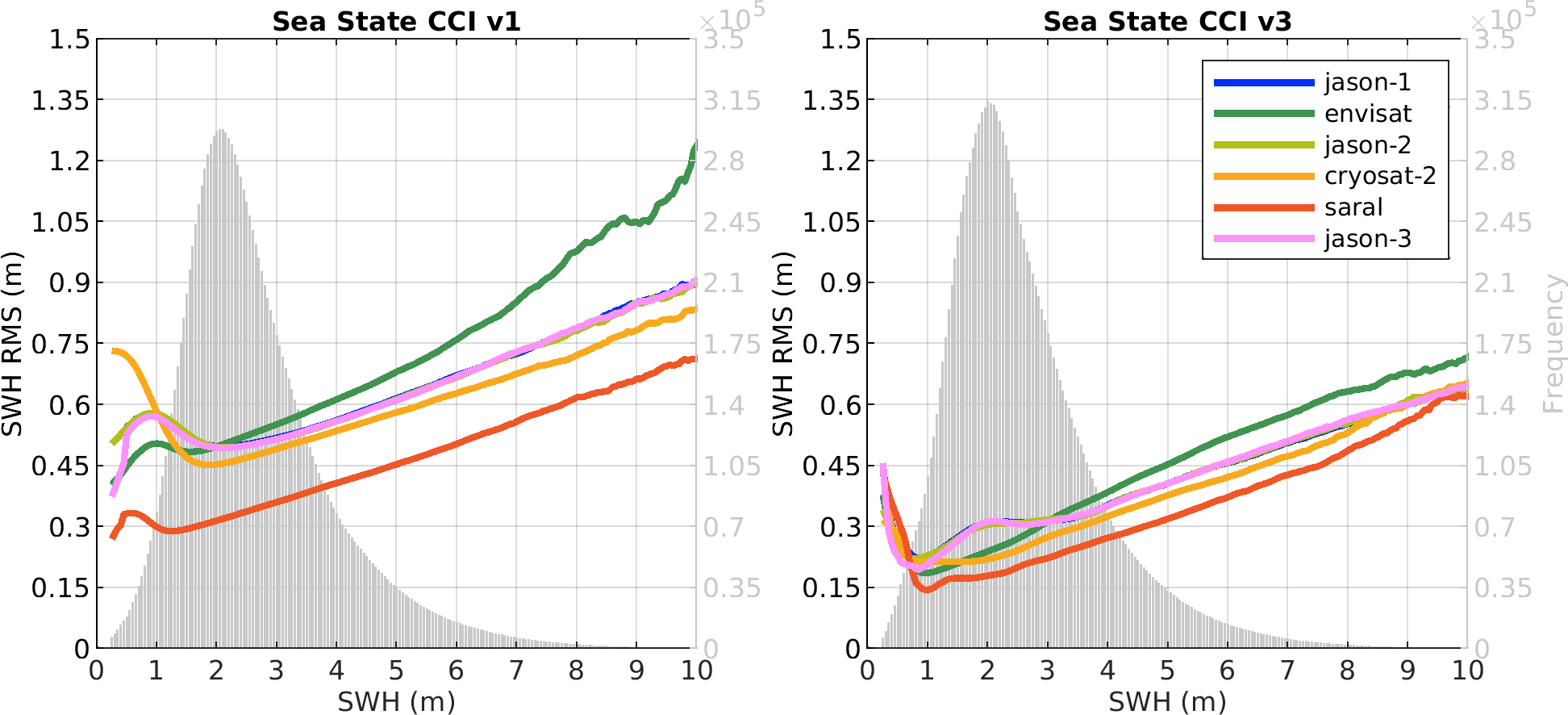}
  \caption{SWH RMS as a function of SWH, estimated from 30 days of data for missions Jason-1, Envisat, Jason-2, Cryosat-2, Saral and Jason-3. Data are from the Sea State CCI dataset version 1 (left) and version 3 (right).}
  \label{fig:valid_comp_swh_rms_vs_swh}
\end{figure}

The root mean square deviation of the 20Hz SWH measurements (SWH RMS) is a standard parameter in altimeter GDRs. It provides a useful metric to estimate along-track noise levels and detect erroneous SWH measurements associated to land contamination, strong rain attenuation, or surface slicks. This parameter is therefore routinely used in altimeter data editing schemes. In Figure \ref{fig:valid_comp_swh_rms_vs_swh}, we show the SWH RMS as a function of SWH, for the Sea State CCI version1 (left) and version 3 (right) dataset. The general trend, independent of the retracker, is a steady increase of the SWH RMS with SWH, for SWH larger than 1-2m. At low SWH, nonlinear behavior is observed, which depends on the waveform retracker/CCI product version. In version1, SWH RMS for the various altimeters associated to SWH between 0-2m present a bell-shaped curve with a maximum around 0.5-1m, while in version3, SWH RMS steadily decrease for SWH comprised between 0-1m. These differences can be explained by the treatment of negative values in the 20Hz to 1Hz compression scheme implemented in version3 (SWH values smaller than -0.5m are flagged as outliers and excluded). Overall, we note a clear reduction of the noise level in version3 dataset, with SWH RMS reduced up to 70\% over the 0-10m SWH range. Moreover, the noise levels are much more similar amongst missions in version3 than in version1. Indeed, the SWH RMS curves are more spread out in v1 with values ranging from 0.3m (Saral) to 1.2m, compared to 0.15-0.75m in version3. 

\begin{figure}[!htb]
  \centering
  \includegraphics[width=\textwidth]{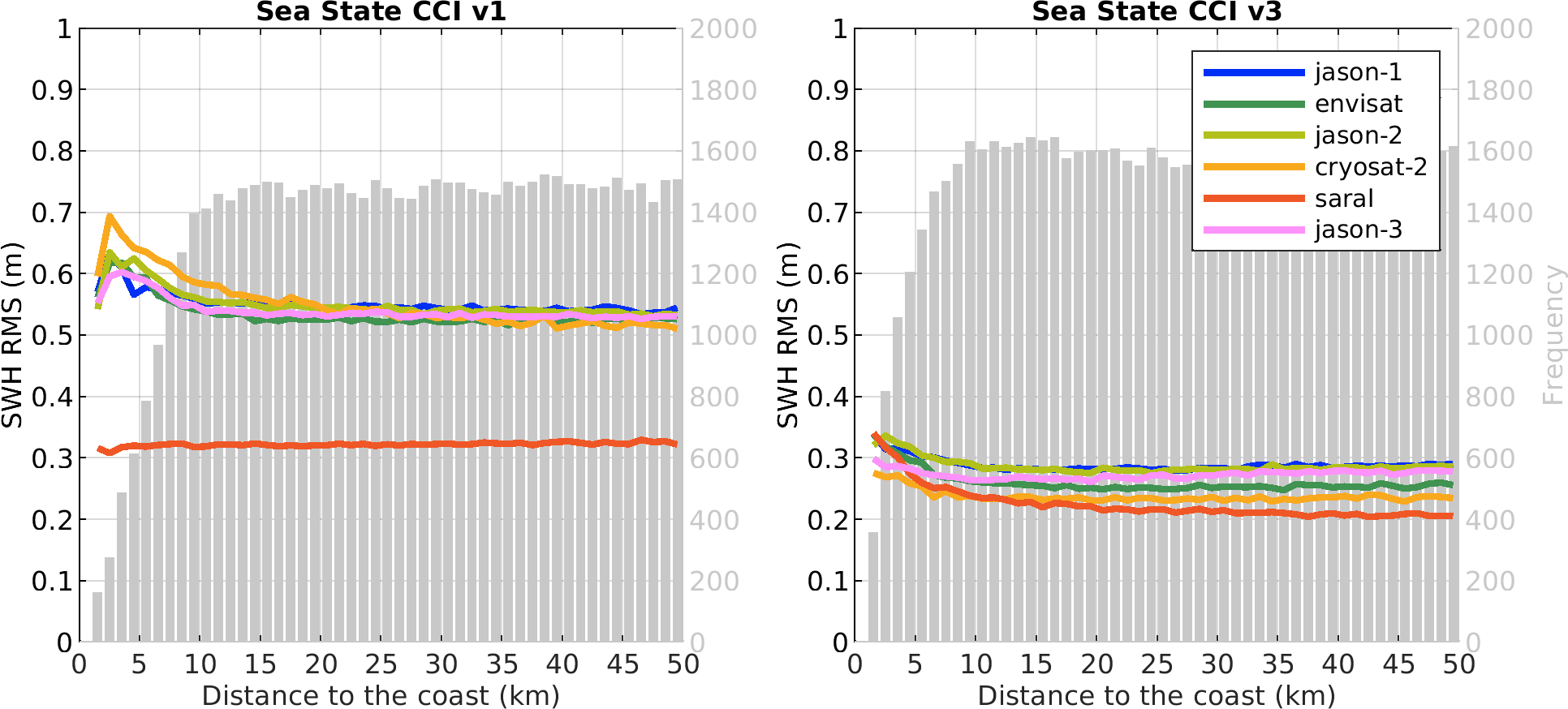}
  \caption{SWH RMS as a function of the distance to the coast, estimated from 30 days of data for missions Jason-1, Envisat, Jason-2, Cryosat-2, Saral and Jason-3. Data are from the Sea State CCI dataset version 1 (left) and version 3 (right).}
  \label{fig:valid_comp_swh_rms_vs_d2c}
\end{figure}

Looking at the distribution of the SWH RMS as a function of the distance to the coast (Figure \ref{fig:valid_comp_swh_rms_vs_d2c}), we note that this parameter remains fairly constant up to 10km to the coast and then it increases. Several differences can be observed between the Sea State CCI version 1 and version 3 dataset. First, as shown before, the noise level in the version1 dataset is clearly higher than in the version 3 dataset. Also, in version1 the SWH RMS presents a maximum around 2-3km to the coast but lower values at 1-2km to the coast, whereas in version3, the maximum SWH RMS is systematically obtained at the closest bin to the coast. As for the previous analysis, this is likely the result of the treatment of negative values in the 20Hz to 1Hz compression scheme, which will impact the very low SWH values predominant in sheltered areas close to the coast. It is interesting to note the quasi-steady noise level all the way to the coast in Saral version1 data, while it rapidly increases between 5 to 0 km in version3 data. We assume that the lack of improvement of Saral between v1 and v3 in the 0-5 km band is an artifact of the WHALES retracker algorithm, which failed to retrieve accurate SWH from specific 40Hz Saral waveforms. This remaining issue has been related to the definition of the leading edge, a similar example related to the largest wave heights records and a possible solution is discussed in section \ref{phenomenal_seas}.

Finally, if we look at the distribution of the number of valid data (grey bars in Figure \ref{fig:valid_comp_swh_rms_vs_swh} and Figure \ref{fig:valid_comp_swh_rms_vs_d2c}) we clearly see that the version3 dataset contains a higher number of valid measurements, whatever the sea state conditions and distance the coast. This increase reaches 15\% at 10km from the coast and 30\% at 5km from the coast.

\subsection{Comparisons with coastal in situ data}
\label{impact_insitu}
\begin{figure}[!htb]
  \centering
  \includegraphics[width=\textwidth]{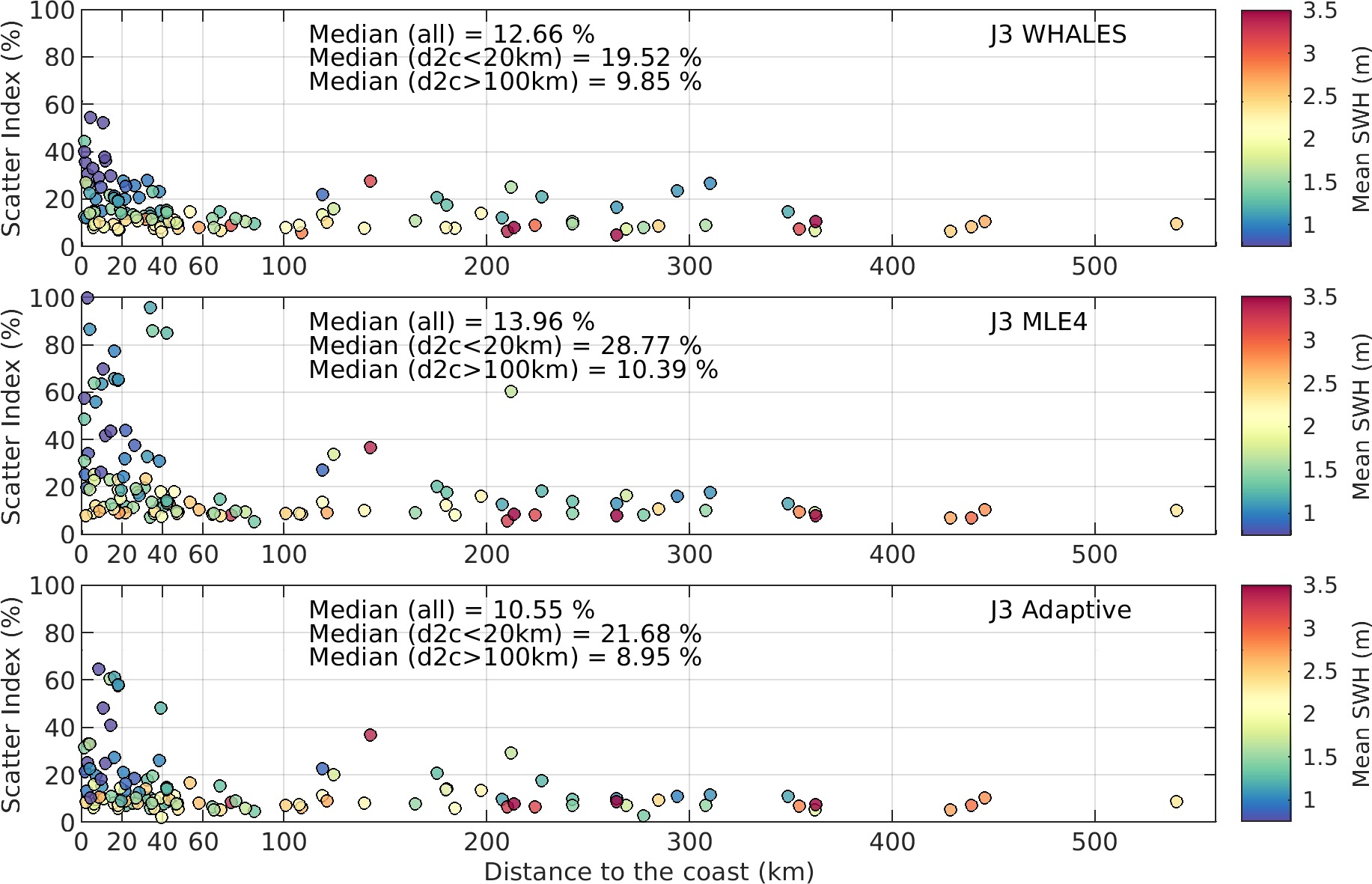}
  \caption{Distribution of the altimeter-insitu scatter index for 122 buoys as a function of the buoy distance to the coast. Colors indicate the mean buoy SWH. Altimeter SWH data are from Jason-3 mission, estimated with WHALES (top panel), MLE4 (middle panel) and Adaptive (bottom panel) retrackers.}
  \label{fig:valid_coastal insitu}
\end{figure}

In order to assess the performance of the WHALES retracker, we compared Jason-3 measurements to in situ measurements acquired by moored wave buoys. We considered a total of 122 wave buoys located within 20 km from Jason-3 ground tracks and at a 1-km minimum distance to the coast. Overall, the buoys are located between 1 and 600 km from the coast, with 60\% located within 50 km from the coast. The buoy data were retrieved from the CMEMS In Situ Thematic Assembly Center (http://www.marineinsitu.eu/). In order to ensure that only good quality measurements are selected, a number of tests were applied and the following buoy SWH data were rejected:
\begin{itemize}
 \item SWH buoy measurements having a constant SWH value over more than 48 hr;
\item SWH buoy measurements with unrealistic position change within a short period of time;
\item SWH, time or latitude/longitude buoy measurements flagged as invalid by native (CMEMS) quality flags ;
\item SWH buoy measurements provided with a numerical precision equal or larger than 0.5 meters;
\item SWH buoy measurements out of the 0-30 meter range;
\end{itemize}
Matchups between altimeter and in situ platform measurements were computed within 20-km distance and 1-h time window. For each satellite pass over a buoy, only the nearest 1Hz altimeter measurement was retained in order to avoid along-track averaging that would inevitably be impacted by the buoy proximity to the coast. Jason-3 SWH records were produced by WHALES together with records from the two retrackers available in the Jason-3 GDRs (version F): MLE-4 \citep{thibaut2010} and Adaptive \citep{tourain2021} retrackers. For the WHALES data we applied the quality flag information from the Sea State CCI version3 dataset. For the MLE-4 and Adaptive data we applied the native quality flags (\textit{swh*compression\_qual} in the AVISO Jason-3 GDRs version F). In order to reduce the impact of varying data editing methodology (more or less stringent) on the final results, we also applied the WHALES (Sea State CCI version3) quality flags to the two other dataset. Figure \ref{fig:valid_coastal insitu} shows the distribution of the scatter index (SI) between altimeter and buoy data, for each wave buoy, as a function of the buoy distance to the coast. Overall, we note a clear reduction of the performance for buoys located within 50km from the coast. This is clearly linked to the calmer sea state conditions encountered in coastal sheltered bays, which are badly resolved by conventional altimetry, corroborating results from \citet{dodet_impact_2025}. Indeed, the large SI values obtained for buoys located within 50km from the coast are mostly associated to low SWH values (as indicated by the color of the circles). Considering the median SI over all buoys, we see that the MLE-4 retracker is showing the worst performance, followed by WHALES and Adaptive. However, if we restrict the analysis to the buoys located within 20km from the coast, we see that WHALES is giving the best performance, followed by Adaptive and MLE-4. These results confirm the excellent performance of WHALES, particularly in the coastal zone, where it outperforms any other retracking algorithms, as previously shown by \cite{schlembach2020}.

%\section{Resolution and precision}
\section{Discussion: the choice of the cost function in the WHALES retracker}
\label{weights_impact}
We now analyze the impact of the design choices of the WHALES retracker cost function on the estimation of the SWH using simulated waveforms. Indeed, one of the choices in designing a retracker is deciding which cost function to use in the fitting process, i.e., when comparing the residuals between the reconstructed waveform and the measured one. The most commonly used schemes, including WHALES, aim to estimate three parameters (epoch, SWH, and amplitude of the waveform, which is related to the backscatter coefficient). The two typical choices for cost functions adopted in the literature are least squares estimation (LS3) and maximum likelihood estimation (ML3). LS3 is the solution adopted in the standard retrackers of most altimetry missions (in their low-resolution mode), although it is misleadingly called “MLE3” for historical reasons. Our interest is to understand the impact of the weights on how the retracker estimates the SWH over a realistic wave field.

\begin{figure*}[!htb]
\centerline{\includegraphics[width=\linewidth]{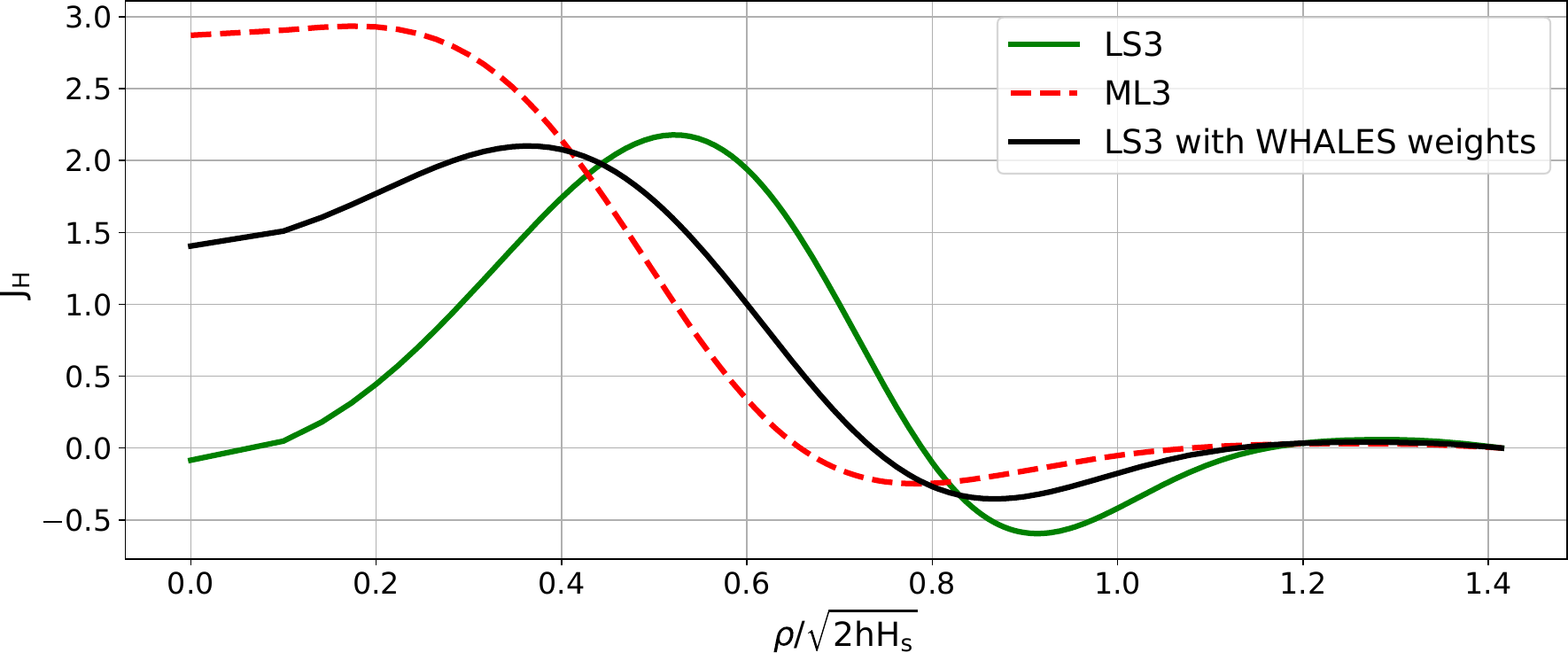}}
\caption{Influence of a wave height perturbation on the retracked wave height, as a function of the distance from nadir $\rho$.  using a 3-parameter least square fit (LS3) with constant or WHALES weights and  a 3-parameter maximum likelihood as discribed in \cite{DeCarlo&Ardhuin2024}.  The perturbed wave height  is $H_s'=Hs(1+\Delta)$ and it is applied to an area $A$, such that  the normalized waveform pertubation is $a=4 \Delta A/\pi \rho_c^2 = 0.02$.
}
\label{fig:Jfunctions}
\end{figure*}

We can return to the assumptions on which the Brown model is based on, in particular the spatially homogeneous wave heights, to investigate the impact of gradients in the local wave height that occur even when the $H_s$ is constant, due to the presence of wave groups \citep{DeCarlo&Ardhuin2024}. When retracking uses a least squares cost function, the fitted waveform parameters, i.e. the wave height and epoch, are well approximated by a linear combinations of parameter perturbations associated to local wave height perturbations.  Figure \ref{fig:Jfunctions} shows the typical shapes of the sensitivity of different retracker cost functions $J_H$ to localized perturbations in wave heights. With non-zero values at nadir, and a weak negative lobe for distances above 0.75 times the Chelton radius $\rho_c = \sqrt{2 h H_s}$, the WHALES weights give intermediate results between the ML3  and LS3 (often wrongly called "MLE3") cost functions. We note that the the lower magnitude of $J_H$ for WHALES weights compared to ML3 should lower the standard deviation of retracked SWH values for very large wave height, for which wave group effects are typically larger than speckle noise effects. 

%%%%%%%%%%%%%%%%%%%%%%%%%%%%%%%%%%%%%%%%%%%%%%%%%%%%%%%%%%%%%%%%%%%%%%%%%%%%%%%%%%
%%%%%%%%%%%%%%%%%%%%%%%%%%%%%%%%%%%%%%%%%%%%%%%%%%%%%%%%%%%%%%%%%%%%%%%%%%%%%%%%%%

\begin{figure*}[!htb]
\centerline{\includegraphics[width=\linewidth]{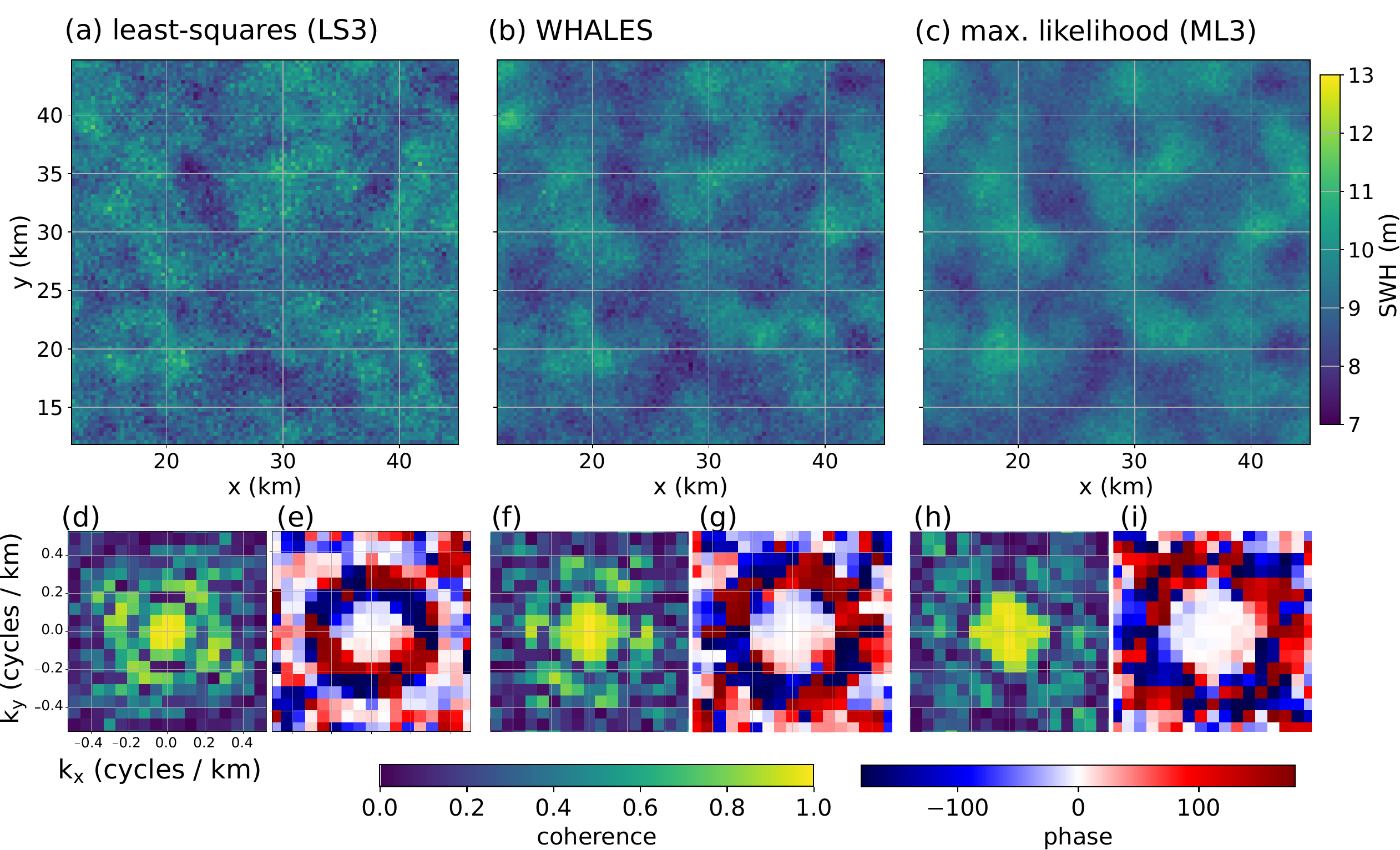}}
\caption{Example of simulated maps of SWH estimated from the same waveforms using 3 different retrackers: (a,d,e) a 3-parameter least square fit (LS3), (b,f,g) WHALES, (c,h,i) a 3-parameter maximum likelihood as discribed in \cite{DeCarlo&Ardhuin2024}. The lower panels (d)--(i) show the coherence and phase of the retracked SWH patterns compared to the Gaussian filtered ($\sigma = 1.6$~km) map of local wave heights.
}
\label{fig:SWHmaps}
\end{figure*}

The practical impact of the retracker cost function is further illustrated in 
Figure \ref{fig:SWHmaps} with simulated SWH estimates using the same sea surface, defined by a spatially uninform spectrum and constant significant wave height $H_s=9.3$~m. This case corresponds to the narrow swell spectrum discussed in  \cite{DeCarlo&Ardhuin2024}, here simulated for the altitude of Jason satellite, hence with a larger value of the radius $\rho_c$ compared to the CFOSAT configuration used in that paper. 

%%%%%%%%%%%%%%%%%%%%%%%%%%%%%%%%%%%%%%%%%%%%%%%%%%%%%%%%%%%%%%%%%%%%%%%%%%%%%%%%%%
%%%%%%%%%%%%%%%%%%%%%%%%%%%%%%%%%%%%%%%%%%%%%%%%%%%%%%%%%%%%%%%%%%%%%%%%%%%%%%%%%%
 The figure clearly shows the intermediate influence of speckle noise (small scale uncorrelated noise) when using the WHALES retracker, with speckle noise most reduced in an adaptive-like ML3 retracker. To quantify the effectively resolved scales in the wave field, one can look at the cross-spectra between the retracked SWH maps and the local wave height maps of the input surface. In Figure Figure \ref{fig:SWHmaps}(d)-(i) we have used cross-spectra estimated with 8 degrees of freedom. In these simulations the Chelton radius is 3.2~km. These are consistent with the shapes of the $J_H$ function: the LS3 retracker has a zero-crossing of $J_H$ at a distance of 0.73$\rho_c=2.3$~km, that corresponds to change in sign of the cross spectra at a wavelength of 4.6~km (the coherence goes to zero and the phase flips from 0 to 180 degrees).  That scale is shorter for WHALES, about 4.2~km,  and even shorter for ML3 at about 3.3 km. However, given the shapes of the $J_H$ function, it is not possible to summarize the filtering effect by a single number. We recall that this filtering effect scales with $\rho_c = \sqrt{2 h H_s}$ and is thus much less important for usual wave heights around 2~m, in which case the speckle is generally the dominant source of noise in SWH estimates. 

\section{Discussion: possible adjustments and evolution of WHALES}
\label{discussion_adjustments}
\begin{figure*}[!htb]
\centerline{\includegraphics[width=\linewidth]{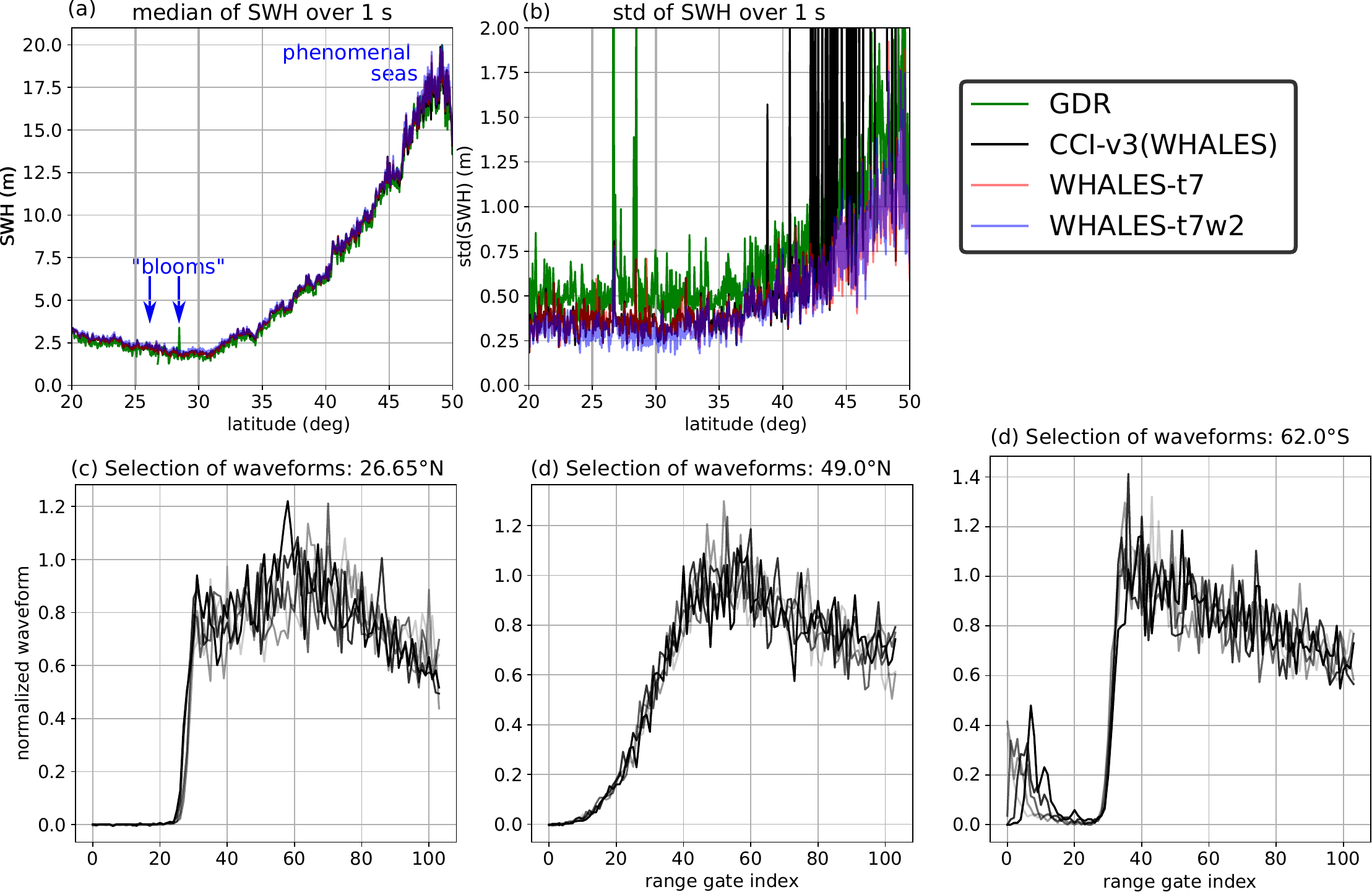}}
\caption{
Wave heights and selected waveforms along Jason-2 cycle 096 track 139 (from GDR D). 
(a,b) median and standard deviation of SWH for each group of 20 waveforms with different retracker solutions. This track has some specific challenges for retrackers, including so-called $\sigma_0$-blooms, and very large wave heights. 5 consecutive waveforms are shown .
%Samples A--E were chosen to illustrate some of the wave patterns found in SWOT data and are all generated by the same storm.
}
\label{fig:Kirin1}
\end{figure*}

The validation detailed in section 3 clearly shows that the version of WHALES that was used to produce the Sea State CCI version3 dataset is very robust and any change to the algorithm that may benefit some specific cases may have side effects on the more usual conditions.  With this in mind, here we discuss three possible types of waveforms that typically occur less than 1\% of the time, and for which WHALES may be improved. For this illustration we have used one satellite track from Jason-2 (cycle 096, track 139). This track was acquired on February 14, 2011 and contains some of the largest wave heights ever recorded by a satellite altimeter \citep{Hanafin&al.2012}. The median and standard deviations are shown in Fig. \ref{fig:Kirin1}a,b for the northern part of the track. We will also discuss data around an iceberg, along the same track, near 137.25$^\circ$W, 62$^\circ$S.   

On these limited tests we generally found that a theoretical weight shape given by the expected noise applied to the first pass of the retracker (option $w=2$) generally reduced the noise in the estimated SWH values. It is not clear if this comes from a continuous dependency on SWH -- which might be achieved with a refined table -- or the better positioning of the weights relative to the waveform.
%%%%%%%%%%%%%%%%%%%%%%%%%%%%%%%%%%%%%%%%%%%%%%%%%%%%%%%%%%%%%%%%%%%%%%%%%%%%%%%%%%
%%%%%%%%%%%%%%%%%%%%%%%%%%%%%%%%%%%%%%%%%%%%%%%%%%%%%%%%%%%%%%%%%%%%%%%%%%%%%%%%%%

\subsection{low winds and/or $\sigma_0$-blooms}
\begin{figure*}[!htb]
\centerline{\includegraphics[width=\linewidth]{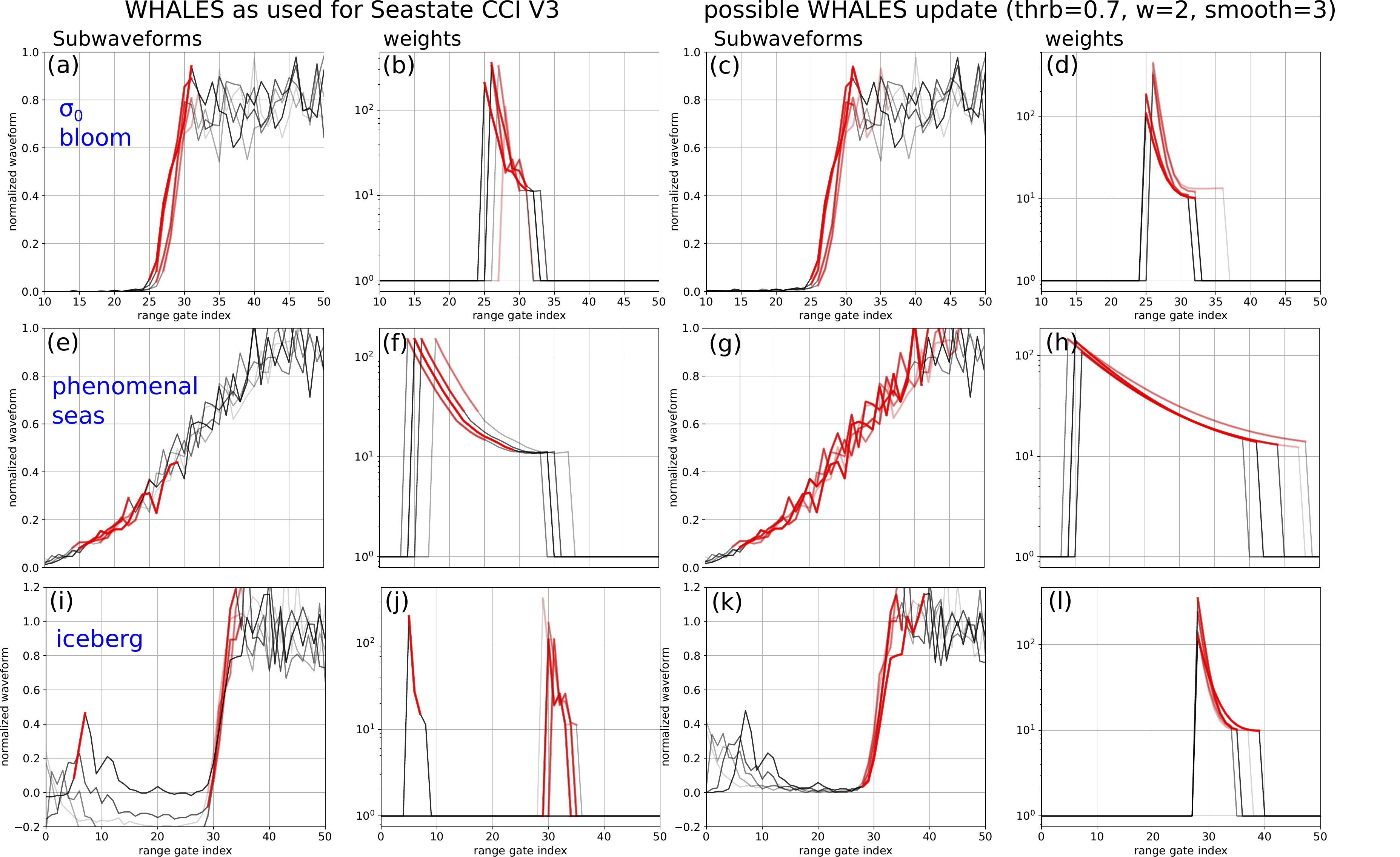}}
\caption{
Subwaveforms and weights for the standard WHALES algorithm and for a possible update.
(a,b,c,d) corresponds to the $\sigma_0$-blooms case shown in Fig. \ref{fig:Kirin1}.c, 
(e,f,g,h) corresponds to the phenomenal sea case shown in Fig. \ref{fig:Kirin1}.f, 
(i,j,k,l) corresponds to the iceberg case shown in Fig. \ref{fig:Kirin1}.e. 
}
\label{fig:Kirin2}
\end{figure*}

A relatively frequent phenomenon at very low wind speeds, which affects about 5\% of ku-band altimeter ocean data \citep{Tournadre&al.2006} is the occurence of patches of high radar backscatter regions, related to the patchiness of the wind speed often combined with the chemical properties of the sea surface where slicks can damp short gravity waves with a strong impact on the mean square slope \citep{Cox&Munk1954} and hence the estimated $\sigma_0$ \citep{Vandemark&al.2004}. In such non-uniform conditions the theoretical Brown model is no longer valid and retracking by standard methods can produce unrealistic values of the estimated parameters. Very often the effect is limited and the result will appear like "noise" in the estimated parameters \citep{Dibarboure&al.2014}, and their impact in applications has been mitigated by editing and averaging the data from the native 20 or 40 Hz resolution to 1 Hz. However, if one wants to interpret the variability of SWH estimates, for example for determining a measurement uncertainty, these effects are an important contributions, as shown in Fig. \ref{fig:Kirin1}.b, and they are due here to the presence of large off-nadir NRCS values as represented in green in Fig. \ref{fig:schematic}.a, which leads to the waveforms shown in Fig. \ref{fig:Kirin1}.c. Compared to the GDR dataset, WHALES removes most of the outliers. This is due to the use of a sub-waveform as shown in Fig. \ref{fig:Kirin2}.a. 

%%%%%%%%%%%%%%%%%%%%%%%%%%%%%%%%%%%%%%%%%%%%%%%%%%%%%%%%%%%%%%%%%%%%%%%%%%%%%%%%%%
%%%%%%%%%%%%%%%%%%%%%%%%%%%%%%%%%%%%%%%%%%%%%%%%%%%%%%%%%%%%%%%%%%%%%%%%%%%%%%%%%%

The few remaining outliers possibly correspond to situations in which the slicks are close to nadir and thus produce peaks that cannot be separated from the leading edge. There is not much that can be done in this case, and the editing and averaging step are still needed to produce adequate 1 Hz data. In this case the standard deviation of the estimated SWH cannot be interpreted as an uncertainty of the SWH estimate.  

\subsection{Phenomenal seas}
\label{phenomenal_seas}
Although the largest wave heights have often been considered with doubt because of the lack of in situ validation, altimeters are in fact the only instruments that routinely report SHW values over 15~m, and these measurements are both self-consistent and consistent with remote swell observations of very long period waves and numerical models \citep{Hanafin&al.2012,Ardhuin&DeCarlo2025}. There is thus no reason to doubt the very large values a priori. On the contrary, a dedicated analysis of these cases should be encouraged given the unique capabilities of altimeters for measuring "phenomenal seas", defined as conditions with $H_s > 14$~m. Minor changes in all algorithms, including WHALES, can thus benefit to the analysis of phenomenal seas. 
In our case, the largest SWH value in the GDR data is at 49$^\circ$N, and the standard processing in both the GDR and WHALES produce a large number of spurious SWH values. Here we found that the main problem with WHALES was its very conservative definition of the leading edge, excluding range gates as soon as the power starts to decrease, as shown in Fig \ref{fig:Kirin2}.e. Even though the leading edge is extended (grey part of the weight curve in  Fig. \ref{fig:Kirin2}.f), it does not match what we may think is the full leading edge. Several corrections can be added to fix this problem. One approach may be to smooth the waveforms to make sure that oscillations that occur on this type of very broad leading edge may not be detected as the end of the leading edge. The difficulty in that case is when and where to decide to apply the smoothing which may have an impact for lower sea states. Another approach illustrated here is to impose one additional criterion: the end of the leading edge should have a normalized waveform with a high enough power (0.5 to 0.8 may work, here we found that 0.7 was a good compromise). However, that criteria can fail with complex waveforms in coastal areas, and smoothing the waveform before leading edge detection can be a simple practical alternative.

\subsection{Icebergs and thermal noise level}
Another frequent case of non-Brown waveform in the open ocean is associated to the presence of icebergs. In that case, the large ice elevations (50 to 100 m above the sea level) introduces echoes that are particularly challenging to retrack. In some of the cases, these echoes are clearly separated from the leading edge ( Fig \ref{fig:Kirin2}.i) and one can re-define the thermal noise level as the minimum (and not the mean) over the range gates usually occupied by the thermal noise.

\section{Conclusion}
\label{conclusion}
The WHALES retracker described and discussed in this paper was defined as a generalization of the sub-waveform retracker ALES with the intention to optimize the estimation of significant wave heights from nadir altimeters. As already demonstrated by  \cite{schlembach2020} over a limited set of data, it was found to be the best retracker for coastal regions. In the open ocean WHALES had better results than the so-called 'MLE3' and 'MLE4' retrackers used in most GDRs, but a slightly higher noise than the adaptive retracker that has been used for CFOSAT. These features can be explained by the choice of the retracker cost function and the influence of speckle noise and local wave height fluctuations (usually dominated by wave group effects).  

One of the main advantages of the WHALES approach is its versatility in application to all LRM altimetry missions. The current phase of the Sea State CCI project is therefore focused on extending it to those present (Sentinel-6 low rate mode, CFOSAT, SWOT-nadir) and past (ERS1/2) altimetry missions still missing in the archive, in order to guarantee three decades of consistent SWH records for climatic research. While it is important to converge on a single retracking solution that ensures robust performance and the best possible compromise between data quality and data quantity, we must also state in conclusion that the main steps of the WHALES retracker (leading edge identification, adaptive subwaveform, weighted residual analysis) can also be applied to other waveform models. This includes, for example, the use of a waveform model in which the measured point target response is used, as done in \cite{tourain2021}, rather than its approximation through a Gaussian as in the Brown model.

\section*{Data availability}
Version3 and version1 data of the ESA Sea State Climate Change Initiative are freely available under the ESA Climate Data Portal \url{https://climate.esa.int/en/projects/sea-state/}

\section*{Code availability}
The source code for the WHALES algorithm is openly available under the GNU Lesser General Public License (LGPL) version 2.1 on GitHub at:
https://github.com/ne62rut/whales/tree/main/src

\section*{Acknowledgements} This research has been funded by the European Space Agency as part of the Sea State CCI project of the Climate Change Initiative (CCI) (contract no. 4000123651/18/I-NB)

\appendix

\section{Additional Figures}

\begin{figure*}[!htb]
\centerline{\includegraphics[width=\linewidth]{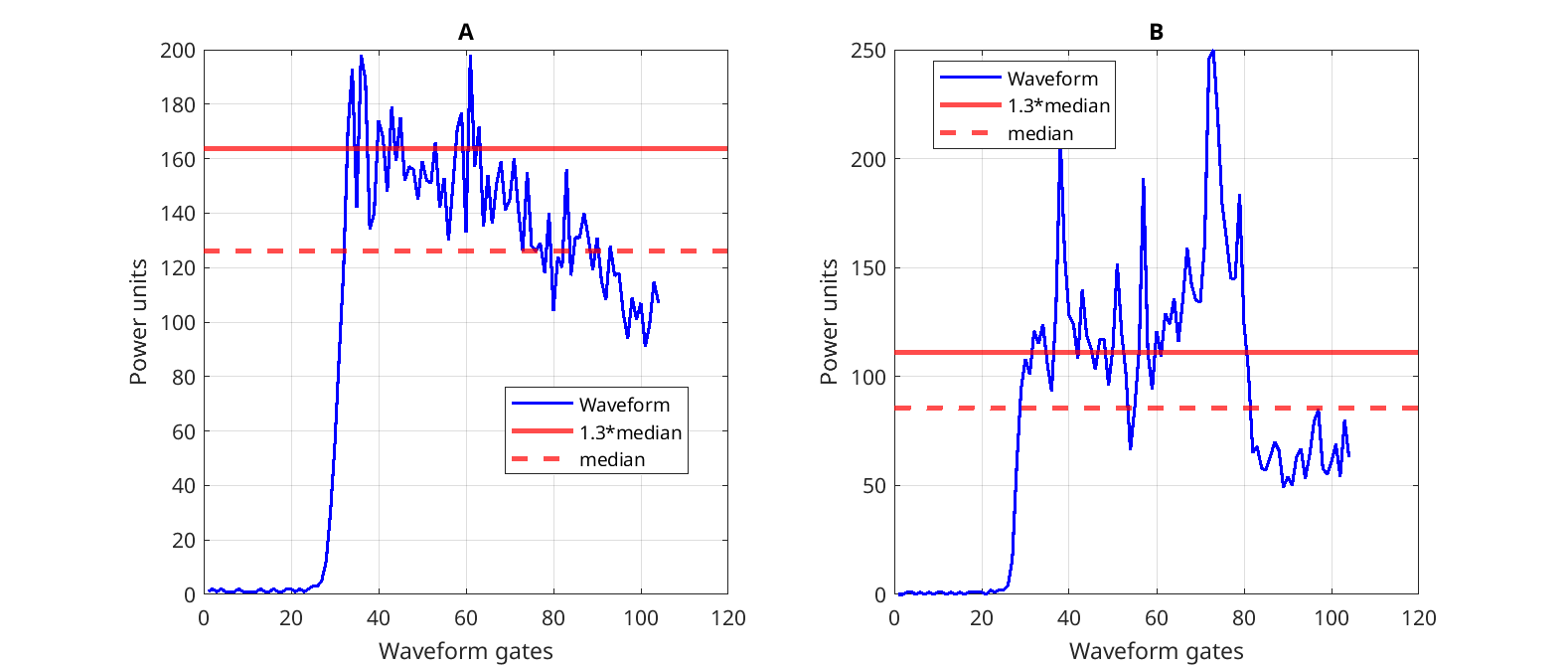}}
\caption{Examples of two waveforms from Jason-2 and their normalization in WHALES. A: an ocean waveform; B: a waveform affected by bright targets in the trailing edge.The red line shows that the chosen normalization factor (1.3 × median of the waveform power) is, in both cases, close to the maximum power of the leading edge.
}
\label{additionalfigures}
\end{figure*}

\bibliographystyle{elsarticle-harv} 

%\bibliography{wave,coastalaltimetryreview}


%\begin{thebibliography}{00}


%\end{thebibliography}

\end{document}